\documentclass[graybox]{svmult}

\usepackage{type1cm}        
\usepackage{makeidx}         
\usepackage{graphicx}        
\usepackage{multicol}        
\usepackage[bottom]{footmisc}

\usepackage{newtxtext}       %
\usepackage{newtxmath}       

\usepackage{cite}
\usepackage{graphbox}

\makeindex             

\newcommand{\bq}{\begin{eqnarray}}
\newcommand{\eq}{\end{eqnarray}}
\newcommand{\eps}{\varepsilon}

\newcommand{\slashoperator}[2]{|_{#2} #1}

\begin{document}

\title*{Iterated integrals related to Feynman integrals associated to elliptic curves}
\titlerunning{Elliptic Integrals}
\author{Stefan Weinzierl}
\institute{Stefan Weinzierl \at PRISMA Cluster of Excellence, Institut f{\"u}r Physik, Johannes Gutenberg-Universit{\"a}t Mainz, D - 55099 Mainz, Germany, \email{weinzierl@uni-mainz.de}
}
\maketitle


\abstract{
This talk reviews Feynman integrals, which are associated to elliptic curves.
The talk will give an introduction into the mathematics behind them,
covering the topics of elliptic curves, elliptic integrals, modular forms and the moduli space
of $n$ marked points on a genus one curve.
The latter will be important, as elliptic Feynman integrals can be expressed as iterated integrals
on the moduli space ${\mathcal M}_{1,n}$,
in same way as Feynman integrals which evaluate to multiple polylogarithms can be expressed 
as iterated integrals on the moduli space ${\mathcal M}_{0,n}$.
With the right language, many methods from the genus zero case carry over to the genus one case.
In particular we will see in specific examples that the differential equation for elliptic Feynman integrals can be cast into an
$\varepsilon$-form.
This allows to systematically obtain a solution order by order in the dimensional regularisation parameter.
}

\section{Introduction}
\label{weinzierl_section_introduction}

In this talk we review Feynman integrals associated to elliptic curves and the mathematics behind them.
It has become common practice to call these Feynman integrals ``elliptic Feynman integrals''.
Elliptic Feynman integrals and closely related integrals in string theory have received considerable attention 
in recent years \cite{Broadhurst:1993mw,Laporta:2004rb,Bailey:2008ib,MullerStach:2011ru,Adams:2013nia,Bloch:2013tra,Remiddi:2013joa,Adams:2014vja,Adams:2015gva,Adams:2015ydq,Bloch:2016izu,Adams:2017ejb,Bogner:2017vim,Adams:2018yfj,Honemann:2018mrb,Bloch:2014qca,Sogaard:2014jla,Tancredi:2015pta,Primo:2016ebd,Remiddi:2016gno,Adams:2016xah,Bonciani:2016qxi,vonManteuffel:2017hms,Adams:2017tga,Ablinger:2017bjx,Primo:2017ipr,Passarino:2017EPJC,Remiddi:2017har,Bourjaily:2017bsb,Hidding:2017jkk,Broedel:2017kkb,Broedel:2017siw,Broedel:2018iwv,Lee:2017qql,Lee:2018ojn,Adams:2018bsn,Adams:2018kez,Broedel:2018qkq,Bourjaily:2018aeq,Besier:2018jen,Mastrolia:2018uzb,Ablinger:2018zwz,Frellesvig:2019kgj,Broedel:2019hyg,Blumlein:2019svg,Broedel:2019tlz,Bogner:2019lfa,Kniehl:2019vwr,Broedel:2019kmn,Abreu:2019fgk,Duhr:2019rrs,2019arXiv190811815L,Walden:2020odh,Weinzierl:2020fyx,Campert:2020yur,Bezuglov:2020ywm,Broedel:2014vla,Broedel:2015hia,Broedel:2017jdo,DHoker:2015wxz,Hohenegger:2017kqy,Broedel:2018izr}.

We call a Feynman integral elliptic, if it can be expressed as a linear combination of iterated integrals 
on a covering space of the moduli space ${\mathcal M}_{1,n}$ of
a genus one curve with $n$ marked points with integrands having only simple poles.
``Ordinary'' Feynman integrals, which evaluate to multiple polylogarithms, can
be expressed as a linear combination of iterated integrals 
on a covering space of the moduli space ${\mathcal M}_{0,n}$ of
a genus zero curve with $n$ marked points, again with integrands having only simple poles.

This definition already uses some mathematical terminology, which we explain in the sequel.
As a rough guide, elliptic Feynman integrals are the next-to-easiest Feynman integrals, with Feynman integrals
evaluating to multiple polylogarithms being the easiest Feynman integrals.
Of course, there are also more complicated Feynman integrals 
beyond these two categories \cite{Brown:2010a,Bourjaily:2018yfy,Bourjaily:2019hmc,Klemm:2019dbm,Bonisch:2020qmm}.
These more complicated integrals are not the topic of this talk.

\section{Background from mathematics}
\label{weinzierl_section_background}

We review the background from mathematics.
The material presented in this section is probably well-known to mathematicians.
It might help physicists as a starting guide into this topic.
Textbooks on elliptic curves are Du Val \cite{Du_Val} and Silverman \cite{Silverman},
textbooks on modular forms are Stein \cite{Stein}, Miyake \cite{Miyake} and Diamond and Shurman \cite{Diamond}.

We start with the definition of an algebraic curve. As ground field we take the complex numbers $\mathbb{C}$.
An algebraic curve in $\mathbb{C}^2$ is defined by the zero set of a polynomial $P(x,y)$ in two variables $x$ and $y$:
\bq
 P\left(x,y\right) & = & 0
\eq
It is more common to consider algebraic curves not in the affine space $\mathbb{C}^2$, but in the projective space
$\mathbb{CP}^2$.
Let $[x:y:z]$ be homogeneous coordinates of $\mathbb{CP}^2$.
An algebraic curve in $\mathbb{CP}^2$ is defined by the zero set of a homogeneous polynomial $P(x,y,z)$ in the three variables $x$, $y$ and $z$:
\bq
\label{weinzierl:def_algebraic_curve}
 P\left(x,y,z\right) & = & 0
\eq
The requirement that $P(x,y,z)$ is a homogeneous polynomial is necessary to have a well-defined zero set on $\mathbb{CP}^2$.

We usually work in the chart $z=1$. In this chart eq.~(\ref{weinzierl:def_algebraic_curve}) reduces to
\bq
 P\left(x,y,1\right) & = & 0.
\eq
If $d$ is the degree of the polynomial $P(x,y,z)$, the arithmetic genus of the algebraic curve is given by
\bq
\label{weinzierl:def_genus}
 g & = & \frac{1}{2} \left(d-1\right) \left(d-2\right).
\eq
For a smooth curve the arithmetic genus equals the geometric genus, 
therefore just using ``genus'' is unambiguous in the smooth case.
Let's look at an example: The equation
\bq
\label{weinzierl:example_elliptic_curve}
 y^2 z - x^3 - x z^2 & = & 0
\eq
defines a smooth algebraic curve of genus $1$.

Let us now turn to elliptic curves:
An elliptic curve over $\mathbb{C}$ is a smooth algebraic curve in $\mathbb{CP}^2$ of genus one with one marked point.
It is common practice to work in the chart $z=1$ and to take as the marked point the ``point at infinity''.
Eq.~(\ref{weinzierl:example_elliptic_curve}) reads in the chart $z=1$
\bq
 y^2 - x^3 - x & = & 0,
\eq
The point at infinity, which is not contained in this chart, is given by $[x:y:z]=[0:1:0]$.

Over the complex numbers ${\mathbb C}$ any elliptic curve can be cast into 
the Weierstrass normal form.
In the chart $z=1$ the Weierstrass normal form reads
\bq
 y^2 & = & 4 x^3 - g_2 x - g_3.
\eq
A second important example is to define an elliptic curve by 
a quartic polynomial in the chart $z=1$:
\bq
\label{weinzierl:example_quartic}
 y^2 & = & \left(x-x_1\right) \left(x-x_2\right) \left(x-x_3\right) \left(x-x_4\right).
\eq
If all roots of the quartic polynomial on the right-hand side are distinct, this defines a smooth elliptic curve.
(The attentive reader may ask, how this squares with the genus formula above.
The answer is that the elliptic curve in $\mathbb{CP}^2$ is not given by the homogenisation
$y^2 z^2 = (x-x_1z)(x-x_2z)(x-x_3z)(x-x_4z)$. The latter curve is singular at infinity.
However, there is a smooth elliptic curve, which in the chart $z=1$ is isomorphic to the affine curve defined by
eq.~(\ref{weinzierl:example_quartic}).
)

As one complex dimension corresponds to two real dimensions, we may consider a smooth algebraic curve 
(i.e. an object of complex dimension one) also as a real surface (i.e. an object of real dimension two).
The latter objects are called Riemann surfaces, as the real surface inherits the structure of a complex manifold.
We may therefore view an elliptic curve either as a complex one-dimensional smooth algebraic curve in $\mathbb{CP}^2$
with one marked point or as a real Riemann surface of genus one with one marked point.
\begin{figure}
\begin{center}
\includegraphics[align=c,scale=1.0]{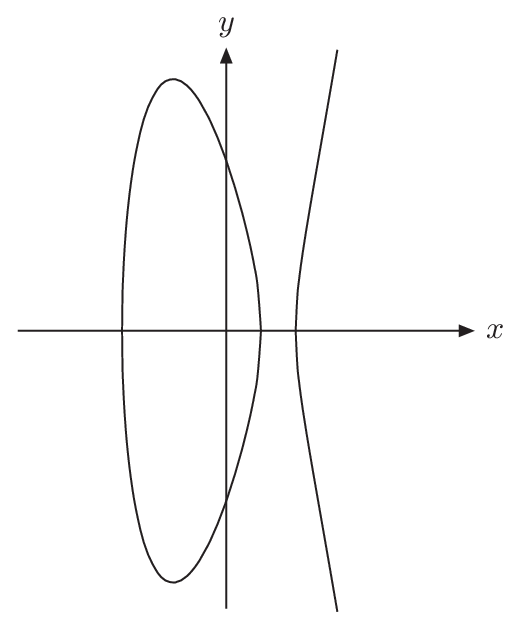}
\hspace*{10mm}
\includegraphics[align=c,scale=1.0]{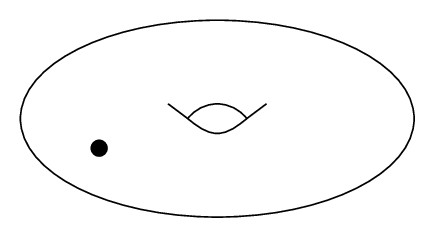}
\end{center}
\caption{
The left picture shows the real part of an elliptic curve in 
the Weierstrass normal form $y^2 = 4 x^3 - g_2 x - g_3$.
The marked point is at infinity.
The right part shows a 
real Riemann surface of genus one with one marked point.
}
\label{weinzierl:fig1}
\end{figure}
This is shown in fig.~\ref{weinzierl:fig1}.

Let us now turn to a second topic: Periodic functions.
We consider a non-constant meromorphic function $f$ of a complex variable $z$.
A period $\omega$ of the function $f$ is a constant such that
for all $z$:
\bq
 f\left(z+\omega\right) & = & f\left(z\right)
\eq
The set of all periods of $f$ forms a lattice, which is either 
\begin{itemize}
\item trivial (i.e. the lattice consists of $\omega=0$ only),
\item a simple lattice, generated by one period $\omega$ : $\Lambda = \left\{ n \omega \; | \; n \in {\mathbb Z} \right\}$,
\item a double lattice, generated by two periods $\omega_1, \omega_2$ with $\mathrm{Im}(\omega_2/\omega_1) \neq 0$ : 
\bq
\Lambda & = & \left\{ n_1 \omega_1 + n_2 \omega_2 \; | \; n_1, n_2 \in {\mathbb Z} \right\}.
\eq
It is common practice to order these two periods such that $\mathrm{Im}(\omega_2/\omega_1) > 0$.
\end{itemize}
There cannot be more possibilities: Assume that there is a third period $\omega_3$, which is not an element of the lattice
$\Lambda$ spanned by $\omega_1$ and $\omega_2$.
In this case we may construct arbitrary small periods as linear combinations of $\omega_1$, $\omega_2$ and $\omega_3$
with integer coefficients.
In the next step one shows that this implies that the derivative of $f(z)$ vanishes at any point $z$, hence $f(z)$ is a constant.
This contradicts our assumption that $f$ is a non-constant function.

An example for a singly periodic function is given by
\bq
 \exp\left(z\right).
\eq
In this case the simple lattice is generated by $\omega = 2 \pi i$.

Double periodic functions are called elliptic functions.
An example for a doubly periodic function is given by Weierstrass's $\wp$-function.
Let $\Lambda$ be the lattice generated by $\omega_1$ and $\omega_2$.
Then
\bq
 \wp\left(z\right)
 & = & 
 \frac{1}{z^2} + \sum\limits_{\omega \in \Lambda \backslash \{0\}} \left( \frac{1}{\left(z+\omega\right)^2} - \frac{1}{\omega^2} \right).
\eq
$\wp(z)$ is periodic with periods $\omega_1$ and $\omega_2$.

Of particular interest are also the corresponding inverse functions. These are in general multivalued functions.
In the case of the exponential function $x=\exp(z)$,
the inverse function is given by
\bq
 z 
 & = & 
 \ln\left(x\right).
\eq
The inverse function to Weierstrass's elliptic function $x=\wp(z)$ is an elliptic integral given by
\bq
 z 
 & = &
 \int\limits_x^\infty \frac{dt}{\sqrt{4t^3-g_2t-g_3}}
\eq
with
\bq
 g_2 = 60 \sum\limits_{\omega \in \Lambda \backslash \{0\}} \frac{1}{\omega^4},
 & \;\;\;\;\;\; &
 g_3 = 140 \sum\limits_{\omega \in \Lambda \backslash \{0\}} \frac{1}{\omega^6}.
\eq
The standard elliptic integrals are classified as
complete or incomplete elliptic integrals and as integrals
of the first, second or third kind.
\begin{table}
\begin{center}
\begin{tabular}{|c|rcl|rcl|}
 \hline
 & \multicolumn{3}{|c|}{complete} & \multicolumn{3}{|c|}{incomplete} \\
 \hline
 first kind &
 $K(x)$ & $=$ &
 $\int\limits_0^1 \frac{dt}{\sqrt{\left(1-t^2\right)\left(1-x^2 t^2 \right)}}$
 &
 $F\left(z,x\right)$
 & $=$ &
 $\int\limits_0^z \frac{dt}{\sqrt{\left(1-t^2\right)\left(1-x^2 t^2 \right)}}$
 \\
 second kind &
 $E(x)$ & $=$ &
 $\int\limits_0^1 dt \frac{\sqrt{1-x^2 t^2}}{\sqrt{1-t^2}}$
 &
 $E\left(z,x\right)$
 & $=$ &
 $\int\limits_0^z dt \frac{\sqrt{1-x^2 t^2}}{\sqrt{1-t^2}}$
 \\
 third kind &
 $\Pi(v,x)$ 
 & $=$ &
 $\int\limits_0^1 \frac{dt}{\left(1-vt^2\right)\sqrt{\left(1-t^2\right)\left(1-x^2 t^2 \right)}}$
 &
 $\Pi\left(v,z,x\right)$
 & $=$ & 
 $\int\limits_0^z \frac{dt}{\left(1-v t^2\right)\sqrt{\left(1-t^2\right)\left(1-x^2 t^2 \right)}}$
 \\
 \hline
\end{tabular}
\end{center}
\caption{
The six standard elliptic integrals. They are classified as complete or incomplete elliptic integrals and as integrals
of the first, second or third kind.
}
\label{weinzierl:table1}
\end{table}
Table~\ref{weinzierl:table1} shows the definition of the six standard elliptic integrals.
The complete elliptic integrals are a special case of the incomplete elliptic integrals and obtained
from the incomplete elliptic integrals by setting the variable $z$ to one.

The classification of elliptic integrals 
as integrals of the first, second or third kind follows the classification of Abelian differentials:
An Abelian differential $f(z)dz$ is called Abelian differential of the first kind, if $f(z)$ is holomorphic.
It is called an Abelian differential of the second kind, if $f(z)$ is meromorphic, but with all residues vanishing.
It is called an Abelian differential of the third kind, if $f(z)$ is meromorphic with non-zero residues.

So far we introduced elliptic curves and elliptic integrals. The link between the two is provided
by the periods of an elliptic curve.
An elliptic curve has one holomorphic differential (i.e. one Abelian differential of the first kind).
If we view the elliptic curve as a genus one Riemann surface (i.e. a torus), we see that there are two independent
\begin{figure}
\begin{center}
\includegraphics[scale=1.0]{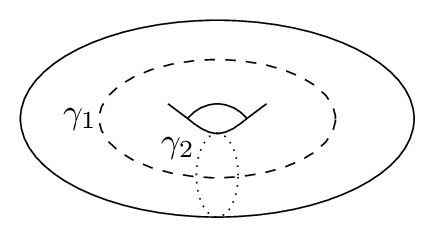}
\end{center}
\caption{
A genus one Riemann surface, where the two independent cycles $\gamma_1$ and $\gamma_2$ are indicated.
}
\label{weinzierl:fig2}
\end{figure}
cycles $\gamma_1$ and $\gamma_2$, as shown in fig.~\ref{weinzierl:fig2}.
A period of an elliptic curve is the integral of the holomorphic differential along a cycle.
As there are two independent cycles, there are two independent periods.
Let's study this for an elliptic curve in the Legendre form
\bq
 y^2 & = & x \left(x-1\right) \left(x-\lambda\right),
\eq
where $\lambda$ is a parameter not equal to $0$, $1$ or infinity.
The periods are
\bq
 \omega_1 = 2 \int\limits_{0}^{\lambda} \frac{dx}{y} = 4 K\left(\sqrt{\lambda}\right),
 & &
 \omega_2 = 2 \int\limits^{\lambda}_{1} \frac{dx}{y} = 4 i K\left(\sqrt{1-\lambda}\right).
\eq
The elliptic curve $y^2 = x (x-1) (x-\lambda)$ depends on a parameter $\lambda$, 
and so do the periods $\omega_1(\lambda)$ and $\omega_2(\lambda)$.
We may now ask:
How do the periods change, if we change $\lambda$?
The variation is governed by a second-order differential equation:
With $t = \sqrt{\lambda}$ we have
\bq
 \left[ t \left(1-t^2\right) \frac{d^2}{dt^2} + \left(1-3t^2\right) \frac{d}{dt} - t \right] \omega_j & = & 0,
 \;\;\;\;\;\; j = 1,2.
\eq
The differential operator
\bq
 t \left(1-t^2\right) \frac{d^2}{dt^2} + \left(1-3t^2\right) \frac{d}{dt} - t
\eq
is called the Picard-Fuchs operator of the elliptic curve $y^2 = x (x-1) (x-\lambda)$.

There is a third possibility to represent an elliptic curve:
We may also represent an elliptic curve as $\mathbb{C}/\Lambda$,
where $\Lambda$ is the double lattice generated by $\omega_1$ and $\omega_2$.
\begin{figure}
\begin{center}
\includegraphics[scale=1.0]{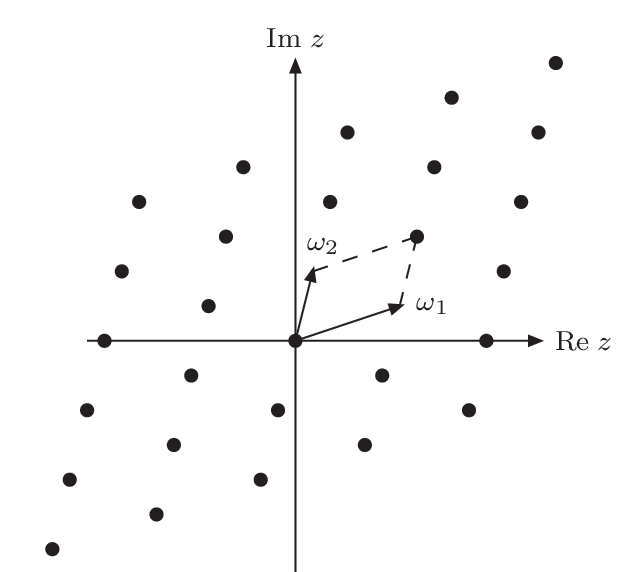}
\end{center}
\caption{
$\mathbb{C}/\Lambda$,
where $\Lambda$ is a double lattice generated by $\omega_1$ and $\omega_2$.
Points inside the fundamental parallelogram correspond to points on the elliptic curve.
A point outside the fundamental parallelogram can always be shifted inside the fundamental parallelogram
through the addition of some lattice vector.
}
\label{weinzierl:fig3}
\end{figure}
This is shown in fig.~\ref{weinzierl:fig3}.
Points, which differ by a lattice vector are considered to be equivalent.
The different equivalence classes are represented by the points inside the fundamental parallelogram,
as shown in fig.~\ref{weinzierl:fig3}.
They correspond to points on the elliptic curve.
Before we go into the details, let us first remark that this is not too surprising:
If we start from the representation of an elliptic curve as a genus one Riemann surface and cut open this surface
along the two cycles $\gamma_1$ and $\gamma_2$ shown in fig.~\ref{weinzierl:fig2}, we obtain
a parallelogram.

Let's now fill in the technical detail: We would like to map a point on an elliptic curve, defined by a
polynomial $P$, to a point in $\mathbb{C}/\Lambda$ and vice versa.
For simplicity we assume that the elliptic curve is given in the Weierstrass normal form
$y^2-4x^3+g_2x+g_3=0$.

We start with the direction from the 
Weierstrass normal form to $\mathbb{C}/\Lambda$:
Given a point $(x,y)$ with $y^2-4x^3+g_2x+g_3=0$ the corresponding point $z \in \mathbb{C}/\Lambda$ is given by
\bq
 z & = &
\int\limits_x^\infty \frac{dt}{\sqrt{4t^3-g_2t-g_3}}.
\eq
Let's now consider the reverse direction from $z \in \mathbb{C}/\Lambda$ to
a point on the curve defined by the Weierstrass normal form.
Given a point $z \in \mathbb{C}/\Lambda$ the corresponding point $(x,y)$ on $y^2-4x^3+g_2x+g_3=0$
is given by
\bq
 \left(x,y\right)
 & = &
 \left( \wp\left(z\right), \wp'\left(z\right) \right).
\eq
$\wp(z)$ denotes Weierstrass's $\wp$-function.

Let us now introduce some additional notation and conventions:
It is common practise to normalise one period to one:
$(\omega_2,\omega_1) \rightarrow (\tau, 1)$, where
\bq
\label{weinzierl:def_tau}
 \tau & = & \frac{\omega_2}{\omega_1}.
\eq
In addition one requires $\mathrm{Im}(\tau) > 0$.
This is always possible: If $\mathrm{Im}(\tau) < 0$ simply  exchange $\omega_1$ and $\omega_2$ and proceed as above.
The possible values of $\tau$ lie therefore in 
the complex upper half-plane, defined by
\bq
 \mathbb{H}
 & = &
 \left\{ \tau \in \mathbb{C} | \mathrm{Im}(\tau) > 0 \right\}.
\eq
Let us now turn to modular transformations:
We have seen that we may represent an elliptic curve as 
$\mathbb{C}/\Lambda$, where $\Lambda$ is a
double lattice generated by $\omega_1$ and $\omega_2$.
\begin{figure}
\begin{center}
\includegraphics[scale=1.0]{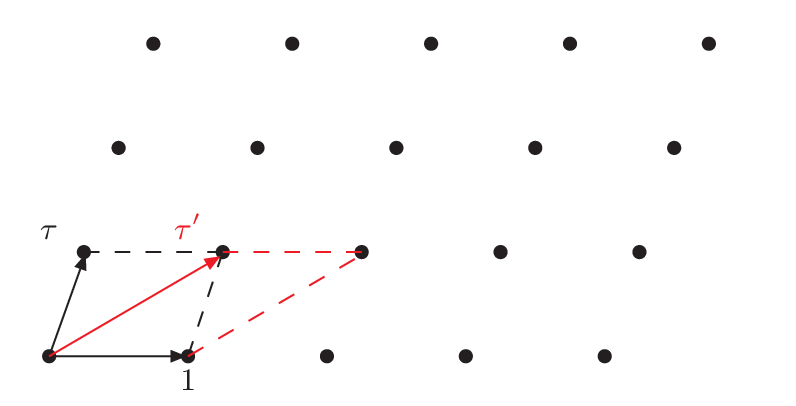}
\end{center}
\caption{
The generators $\tau$ and $1$ generate the same lattice as the generators $\tau'$and $1$.
}
\label{weinzierl:fig4}
\end{figure}
As only the lattice $\Lambda$ matters, but not the specific generators, we may consider a different pair
of periods $(\omega_2',\omega_1')$, which generate the same lattice $\Lambda$.
An example is shown in fig.~\ref{weinzierl:fig4}:
The generators $\tau$ and $1$ generate the same lattice as the generators $\tau'$and $1$.

Let's return to the general case and 
consider a change of basis from the pair of periods $(\omega_2,\omega_1)$ to the pair of periods
$(\omega_2',\omega_1')$.
The new pair of periods $(\omega_2',\omega_1')$ is again a pair of lattice vectors, so it can be written
as
\bq
 \left( \begin{array}{c}
 \omega_2' \\
 \omega_1' \\
 \end{array} \right)
 & = &
 \left( \begin{array}{cc}
 a & b \\
 c & d \\
 \end{array} \right)
 \left( \begin{array}{c}
 \omega_2 \\
 \omega_1 \\
 \end{array} \right),
\eq
with $a, b, c, d \in {\mathbb Z}$.
The transformation should be invertible and $(\omega_2,\omega_1)$ and $(\omega_2',\omega_1')$ should generate the same lattice $\Lambda$.
This implies
\bq
 \left( \begin{array}{cc}
 a & b \\
 c & d \\
 \end{array} \right)
 & \in & \mathrm{SL}_2\left({\mathbb Z}\right).
\eq
In terms of $\tau$ and $\tau'$ we have
\bq
\label{weinzierl:trafo_modular}
 \tau' & = & \frac{a \tau +b}{c \tau +d}.
\eq
A transformation of the form as in eq.~(\ref{weinzierl:trafo_modular}) is called a modular transformation.

We may then look at functions $f(\tau)$, which transform under modular transformations in a particular way.
This leads to modular forms.
A meromorphic function $f: \mathbb{H} \rightarrow \mathbb{C}$ 
is a modular form of modular weight $k$ for $\mathrm{SL}_2(\mathbb{Z})$
if
\begin{enumerate}
\item $f$ transforms under modular transformations as
\bq
\label{weinzierl:trafo_modular_form}
 f\left( \dfrac{a\tau+b}{c\tau+d} \right) = (c\tau+d)^k \cdot f(\tau) 
 \qquad \text{for} \;\; \gamma = \left( \begin{array}{cc}
a & b \\ 
c & d
\end{array} \right) \in \mathrm{SL}_2(\mathbb{Z}),
\eq
\item $f$ is holomorphic on $\mathbb{H}$,
\item $f$ is holomorphic at $i \infty$.
\end{enumerate}
The prefactor $(c\tau+d)^k$ in eq.~(\ref{weinzierl:trafo_modular_form}) is called automorphic factor
and equals
\bq
 (c\tau+d)^k
 & = &
 \left( \frac{\omega_1'}{\omega_1} \right)^k.
\eq
It is convenient to introduce the $\slashoperator{\gamma}{k}$ operator, defined by
\bq
(f \slashoperator{\gamma}{k})(\tau) & = & (c\tau+d)^{-k} \cdot f(\gamma(\tau)).
\eq
With the help of the $\slashoperator{\gamma}{k}$ operator we may rewrite eq.~(\ref{weinzierl:trafo_modular_form})
as
\bq
\label{weinzierl:trafo_modular_form_v2}
 (f \slashoperator{\gamma}{k}) & = & f
 \qquad \text{for} \;\; 
 \gamma 
 \in \mathrm{SL}_2(\mathbb{Z})
\eq
A meromorphic function $f : {\mathbb H} \rightarrow {\mathbb C}$, which only satisfies eq.~(\ref{weinzierl:trafo_modular_form}) (or equivalently only eq.~(\ref{weinzierl:trafo_modular_form_v2})) is called 
weakly modular of weight $k$ for $\mathrm{SL}_2(\mathbb{Z})$.

Apart from $\mathrm{SL}_2({\mathbb Z})$ we may also look at congruence subgroups.
The standard congruence subgroups are defined by
\begin{align}
\Gamma_0(N) &= \left\{ \left( \begin{array}{cc}
a & b \\ 
c & d
\end{array}  \right) \in \mathrm{SL}_2(\mathbb{Z}): c \equiv 0\ \text{mod}\ N \right\},
 \nonumber \\
\Gamma_1(N) &= \left\{ \left( \begin{array}{cc}
a & b \\ 
c & d
\end{array}  \right) \in \mathrm{SL}_2(\mathbb{Z}): a,d \equiv 1\ \text{mod}\ N, \; c \equiv 0\ \text{mod}\ N  \right\},
 \nonumber \\
\Gamma(N) &= \left\{ \left( \begin{array}{cc}
a & b \\ 
c & d
\end{array}  \right) \in \mathrm{SL}_2(\mathbb{Z}): a,d \equiv 1\ \text{mod}\ N, \; b,c \equiv 0\ \text{mod}\ N \right\}.
\nonumber
\end{align}
$\Gamma(N)$ is called the principle congruence subgroup of level $N$.
The principle congruence subgroup $\Gamma(N)$ is a normal subgroup of $\mathrm{SL}_2({\mathbb Z})$.
In general, a subgroup $\Gamma$ of $\mathrm{SL}_2({\mathbb Z})$ is called a congruence subgroup, if there exists an $N$ such that
\bq
 \Gamma\left(N\right) & \subseteq & \Gamma.
\eq
The smallest such $N$ is called the level of the congruence subgroup.

We may now define modular forms for a congruence subgroup $\Gamma$,
by relaxing the transformation law in eq.~(\ref{weinzierl:trafo_modular_form}) to hold only for
modular transformations from the subgroup $\Gamma$, plus holomorphicity on ${\mathbb H}$ and at the cusps.
In detail:
A meromorphic function $f: \mathbb{H} \rightarrow \mathbb{C}$ 
is a modular form of modular weight $k$ for the congruence subgroup $\Gamma$
if
\begin{enumerate}
\item $f$ transforms as
\bq
 (f \slashoperator{\gamma}{k}) & = & f
 \qquad \text{for} \;\; 
 \gamma 
 \in \Gamma,
\eq
\item $f$ is holomorphic on $\mathbb{H}$,
\item $f \slashoperator{\gamma}{k}$ is holomorphic at $i \infty$ for all $\gamma \in \mathrm{SL}_2({\mathbb Z})$.
\end{enumerate}
For a congruence subgroup $\Gamma$ of $\mathrm{SL}_2({\mathbb Z})$ we denote by
${\mathcal M}_k(\Gamma)$ the space of modular forms of weight $k$.
From the inclusions
\begin{align}
\Gamma(N) \subseteq \Gamma_1(N) \subseteq \Gamma_0(N) \subseteq \text{SL}_2(\mathbb{Z})
\end{align}
follow the inclusions
\bq
\label{weinzierl:inclusion}
 \mathcal{M}_k(\mathrm{SL}_2(\mathbb{Z})) \subseteq \mathcal{M}_k(\Gamma_0(N)) \subseteq \mathcal{M}_k(\Gamma_1(N)) \subseteq \mathcal{M}_k(\Gamma(N)).
\eq
For a given $N$, the space $\mathcal{M}_k(\Gamma(N))$ of modular forms of weight $k$ for the principal congruence
subgroup $\Gamma(N)$ is the largest one among the spaces listed in eq.~(\ref{weinzierl:inclusion}).
By definition we have for $f \in \mathcal{M}_k(\Gamma(N))$ and $\gamma \in \Gamma(N)$
\begin{align*}
 f \slashoperator{\gamma}{k} & = f,
 & 
 \gamma & \in \Gamma(N).
\end{align*}
We may ask what happens if we transform by a $\gamma \in \mathrm{SL}_2(\mathbb{Z})$, which does not belong
to the congruence subgroup $\Gamma(N)$.
One may show that in this case we have
\begin{align*}
 f \slashoperator{\gamma}{k} & \in \mathcal{M}_k(\Gamma(N)),
 & 
 \gamma & \in \mathrm{SL}_2(\mathbb{Z}) \backslash \Gamma(N),
\end{align*}
i.e. $f \slashoperator{\gamma}{k}$ is again a modular form of weight $k$ for $\Gamma(N)$, 
although not necessarily identical to $f$.
The proof relies on the fact that $\Gamma(N)$ is a normal subgroup of $\mathrm{SL}_2(\mathbb{Z})$.
This is essential:
If $\Gamma$ is a non-normal congruence subgroup of $\mathrm{SL}_2(\mathbb{Z})$ one has in general
$f \slashoperator{\gamma}{k} \notin \mathcal{M}_k(\Gamma)$.

Modular forms of $\mathrm{SL}_2(\mathbb{Z})$ are invariant under $\tau'=\tau+1$, since
\bq
\left( \begin{array}{cc}
1 & 1 \\ 
0 & 1
\end{array} \right) \in \mathrm{SL}_2(\mathbb{Z}).
\eq
In other words, they are periodic with period $1$: $f(\tau+1)=f(\tau)$.

It is convenient to introduce for
$\tau \in \mathbb{H}$ and $z \in \mathbb{C}$ 
\bq
\label{weinzierl:def_nome_squared}
 \bar{q} = \exp\left(2\pi i \tau\right),
 & &
 \bar{w} = \exp\left(2\pi i z\right).
\eq
(The notation for $z$ and $w$ we will be useful in the next section.)
$\bar{q}=\exp(2\pi i \tau)$
maps the complex upper half-plane $\tau \in \mathbb{H}$ 
to the unit disk $|\bar{q}|<1$.

The maps in eq.~(\ref{weinzierl:def_nome_squared})
trivialises periodicity with period $1$:
\bq
 \bar{q}\left(\tau+1\right) = \bar{q}\left(\tau\right),
 & &
 \bar{w}\left(z+1\right) = \bar{w}\left(z\right).
\eq
On the other hand, 
shifts by $\tau$ correspond to multiplication with $\bar{q}$:
\bq
 \bar{q}\left(\tau+\tau\right) = \bar{q}\left(\tau\right) \cdot \bar{q}\left(\tau\right),
 & &
 \bar{w}\left(z+\tau\right) = \bar{w}\left(z\right) \cdot \bar{q}\left(\tau\right).
\eq

We now introduce iterated integrals of modular forms.
Let $f_1, \dots, f_n$ be modular forms. We set
\bq
\lefteqn{
 I\left(f_1,f_2,...,f_n; \tau \right) 
 = }
 \nonumber \\
 & = &
 \left(2 \pi i \right)^n
 \int\limits_{\tau_0}^{\tau} d\tau_1
 f_1\left(\tau_1\right)
 \int\limits_{\tau_0}^{\tau_1} d\tau_2
 f_2\left(\tau_2\right)
 ...
 \int\limits_{\tau_0}^{\tau_{n-1}} d\tau_n
 f_n\left(\tau_n\right).
\eq
As basepoint we usually take $\tau_0=i\infty$.
Please note that an integral over a modular form is in general not a modular form. 
This is not surprising if we consider the following analogy:
An integral over a rational function is in general not a rational function.

We usually like iterated integrals appearing in solutions of Feynman integrals to have at worst simple poles.
Let's study iterated integrals of modular forms.
As modular forms are holomorphic in the 
complex upper half-plane, there are no poles there.
So the only interesting points are the cusps.
Let's focus on modular forms $f \in \mathcal{M}_k(\mathrm{SL}_2(\mathbb{Z}))$, so the only cusp is at $\tau=i\infty$.
By definition a modular form $f(\tau)$ is holomorphic at the cusp and has a $\bar{q}$-expansion
\bq
 f(\tau) & = & a_0 + a_1 \bar{q} + a_2 \bar{q}^2 + ...,
 \;\;\;\;\;\;\;\;\;\;\;\;
 \bar{q}=\exp(2\pi i \tau).
\eq
The transformation $\bar{q}=\exp(2\pi i \tau)$ transforms the point $\tau=i\infty$ to $\bar{q}=0$ and we have
\bq
 2 \pi i \; f(\tau) d\tau & = & \frac{d\bar{q}}{\bar{q}} \left( a_0 + a_1 \bar{q} + a_2 \bar{q}^2 + ... \right).
\eq
Thus a modular form non-vanishing at the cusp $\tau=i\infty$ has a simple pole at $\bar{q}=0$.


\section{Moduli spaces}
\label{weinzierl_section_moduli_spaces}

This section gives an introduction into moduli spaces.

Let $X$ be a topological space. The configuration space of $n$ ordered points
in $X$ is
\bq
 \mathrm{Conf}_n\left(X\right)
 & = &
 \left\{ \left. \left(x_1,...,x_n\right) \in X^n \right| x_i \neq x_j \; \mbox{for} \; i \neq j \right\}.
\eq
Please note that we require that the points are distinct: $x_i \neq x_j$.
As a simple example consider the configuration space of $2$ ordered points in ${\mathbb R}$:
\bq
 \mathrm{Conf}_2\left({\mathbb R}\right)
 & = &
 \left\{ \left. \left(x_1,x_2\right) \in {\mathbb R}^2 \right| x_1 \neq x_2\right\}.
\eq
$\mathrm{Conf}_2({\mathbb R})$ is the plane ${\mathbb R}^2$ with the diagonal $x_1=x_2$ removed.

As a second example consider the configuration space of $2$ ordered points in the complex projective space
${\mathbb C}{\mathbb P}^1$ (i.e. the Riemann sphere):
\bq
 \mathrm{Conf}_2\left({\mathbb C}{\mathbb P}^1\right)
 & = &
 \left\{ \left. \left(z_1,z_2\right) \in \left({\mathbb C}{\mathbb P}^1\right)^2 \right| z_1 \neq z_2\right\}.
\eq
This is a two-dimensional space.
A M\"obius transformation
\bq
 z' & = & \frac{az+b}{cz+d}
\eq
transforms the Riemann sphere into itself.
These transformations form a group $\mathrm{PSL}\left(2,{\mathbb C}\right)$.
Usually we are not interested in configurations 
\bq
 (z_1,...,z_n) \in \mathrm{Conf}_n\left({\mathbb C}{\mathbb P}^1\right)
 & \mbox{and} &
 (z_1',...,z_n') \in \mathrm{Conf}_n\left({\mathbb C}{\mathbb P}^1\right),
\eq
which differ only by a M\"obius transformation.
This brings us to the definition of the moduli space of the Riemann sphere with $n$ marked points:
\bq 
 {\mathcal M}_{0,n}
 & = &
 \mathrm{Conf}_n\left({\mathbb C}{\mathbb P}^1\right) / \mathrm{PSL}\left(2,{\mathbb C}\right).
\eq
We may use the freedom of M\"obius transformations to fix three points (usually $0$, $1$ and $\infty$).
Therefore
\bq
\label{weinzierl:dim_M_0_n}
 \dim\left( \mathrm{Conf}_n\left({\mathbb C}{\mathbb P}^1\right) \right) & = & n,
 \nonumber \\
 \dim\left({\mathcal M}_{0,n}\right) & = & n-3.
\eq
Let's generalise this:
We are interested in the situation, where the topological space $X$ 
is a smooth algebraic curve $C$ in ${\mathbb C}{\mathbb P}^2$.
This implies that 
there exists a homogeneous polynomial $P(z_1,z_2,z_3)$ such that
\bq
 C & = & \left\{ \left. \left[z_1:z_2:z_3\right] \in {\mathbb C}{\mathbb P}^2 \right| P\left(z_1,z_2,z_3\right) = 0 \right\}. 
\eq
If $d$ is the degree of the polynomial $P(z_1,z_2,z_3)$, the genus $g$ of $C$ is given by
eq.~(\ref{weinzierl:def_genus}).
\begin{figure}
\begin{center}
\includegraphics[scale=0.85]{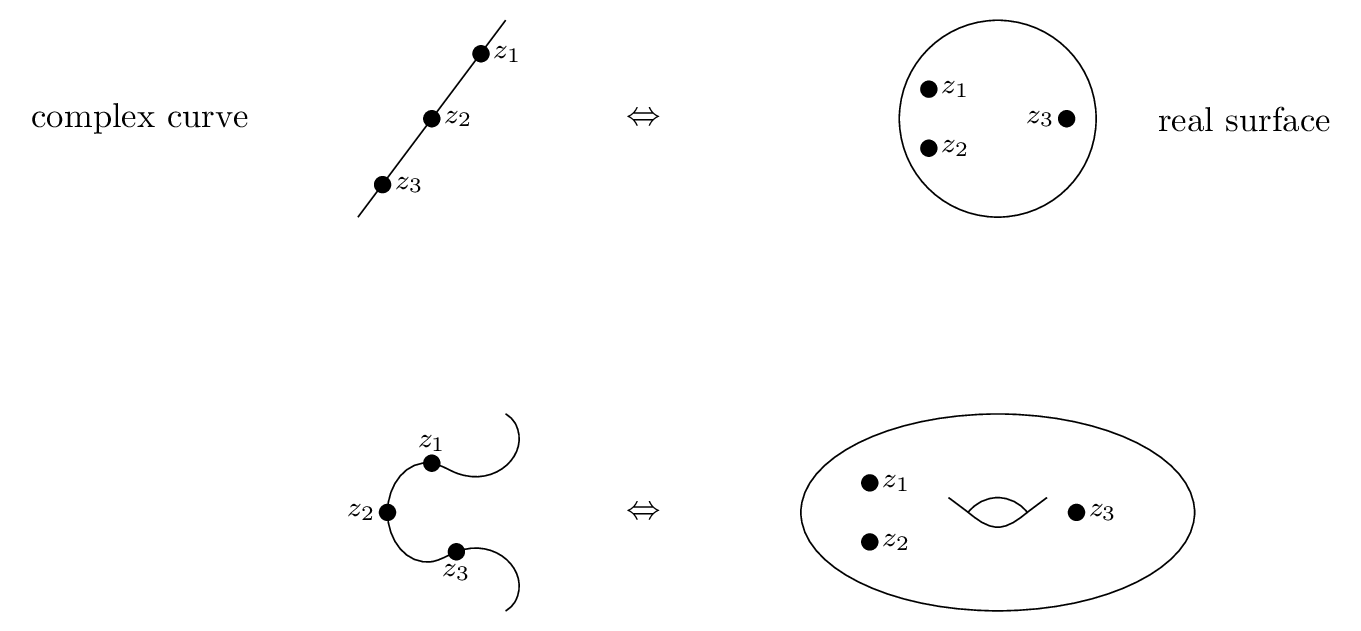}
\end{center}
\caption{
The upper left figure shows a configuration of three marked points on a complex curve of genus zero,
the upper right figure shows the corresponding configuration when the complex curve is viewed as a real
Riemann surface.
The lower figures show the analogous situation for a complex curve of genus one.
}
\label{weinzierl:fig5}
\end{figure}
Alternatively we may view $C$ as a Riemann surface of genus $g$.

Let us now consider a smooth curve $C$ of genus $g$ with $n$ marked points.
Two such curves $(C;z_1,...,z_n)$ and $(C';z_1',...,z_n')$ are isomorphic if there is an isomorphism
\bq
 \phi & : & C \rightarrow C'
 \;\;\;\;\;\;
 \mbox{such that} \;\;
 \phi\left(z_i\right) = z_i'.
\eq
The moduli space
\bq
 {\mathcal M}_{g,n}
\eq
is the space of isomorphism classes of smooth curves of genus $g$ with $n$ marked points.
For $g \ge 1$ the isomorphism classes do not only depend on the positions of the marked points,
but also on the ``shape'' of the curve.
For $g=0$ there is only one ``shape'', the Riemann sphere.

The dimension of ${\mathcal M}_{g,n}$ is
\bq
\label{weinzierl:dim_M_g_n}
 \dim\left({\mathcal M}_{g,n}\right) & = & 3g + n -3,
\eq
for $g=0$ this formula agrees with the previous result in eq.~(\ref{weinzierl:dim_M_0_n}).

In this talk we are mainly interested in the moduli spaces ${\mathcal M}_{0,n}$ and ${\mathcal M}_{1,n}$.
Let us work out natural choices for coordinates on ${\mathcal M}_{0,n}$ and ${\mathcal M}_{1,n}$.
\begin{itemize}
\item
We start with genus $0$.
We have $\dim {\mathcal M}_{0,n} = n - 3$.
As mentioned above, the sphere has a unique shape. 
We may use M\"obius transformations to fix three points, say $z_{n-2}=1$, $z_{n-1}=\infty$, $z_n=0$.
This leaves 
\bq
 (z_1,...,z_{n-3})
\eq
as coordinates on ${\mathcal M}_{0,n}$.
\item
We now turn to genus $1$.
From eq.~(\ref{weinzierl:dim_M_g_n}) we have $\dim {\mathcal M}_{1,n} = n$.
We need one coordinate to describe the shape of the elliptic curve (or the shape of the torus or the shape
of the parallelogram). We may take $\tau$ as defined in eq.~(\ref{weinzierl:def_tau}) for this.
We may use translation to fix one marked point, say $z_n=0$.
This gives
\bq
 (\tau,z_1,...,z_{n-1})
\eq
as coordinates on ${\mathcal M}_{1,n}$.
\end{itemize}
We then consider iterated integrals on ${\mathcal M}_{0,n}$ and ${\mathcal M}_{1,n}$.
In general, iterated integrals are defined as follows:
Let $\omega_1$, ..., $\omega_k$ be differential 1-forms on a manifold $M$
and $\gamma : [0,1] \rightarrow M$ a path.
We write for the pull-back of $\omega_j$ to the interval $[0,1]$
\bq
 f_j\left(\lambda\right) d\lambda & = & \gamma^\ast \omega_j.
\eq
Iterated integral are defined by \cite{Chen}
\bq
 I_{\gamma}\left(\omega_1,...,\omega_k;\lambda\right)
 & = &
 \int\limits_0^{\lambda} d\lambda_1 f_1\left(\lambda_1\right)
 \int\limits_0^{\lambda_1} d\lambda_2 f_2\left(\lambda_2\right)
 ...
 \int\limits_0^{\lambda_{k-1}} d\lambda_k f_k\left(\lambda_k\right).
\eq
Let us now specialise to
iterated integrals on ${\mathcal M}_{0,n}$.
We are interested in differential one-forms, which have only simple poles.
We therefore consider
\bq
 \omega^{\mathrm{mpl}} & = & \frac{dy}{y-z_j}.
\eq
The iterated integrals constructed from these differential one-forms are the 
multiple polylogarithms:
\bq
G(z_1,...,z_k;y) & = & \int\limits_0^y \frac{dy_1}{y_1-z_1}
 \int\limits_0^{y_1} \frac{dy_2}{y_2-z_2} ...
 \int\limits_0^{y_{k-1}} \frac{dy_k}{y_k-z_k},
 \;\;\;\;\;\;
 z_k \neq 0.
\eq
We may slightly enlarge the set of functions by setting
\bq
\label{weinzierl:only_trailing_zeros}
 G(\underbrace{0,\dots,0}_{k};y)
 & = &
 \frac{1}{k!} \ln^k\left(y\right)
\eq
and for $(z_1,z_2,\dots,z_k)\neq (0,0,\dots,0)$
\bq
G(z_1,z_2,...,z_k;y) & = & \int\limits_0^y \frac{dy_1}{y_1-z_1} G(z_2,...,z_k;y_1).
\eq
This allows trailing zeros. We say that the multiple polylogarithm $G(z_1,...,z_k;y)$ has a trailing zero if $z_k=0$.
Using the shuffle product we may convert any multiple polylogarithm with trailing zeros into 
multiple polylogarithm without trailing zeros and powers of $\ln(y)$.

Let's now consider iterated integrals on ${\mathcal M}_{1,n}$.
We recall that we may take $(\tau,z_1,...,z_{n-1})$ as coordinates on ${\mathcal M}_{1,n}$.
We may decompose an arbitrary integration path into pieces along $d\tau$ (with $z_1=\dots =z_{n-1}=\mathrm{const}$) 
and pieces along the $dz_j$'s (with $\tau=\mathrm{const}$).
Thus we obtain two classes of standardised iterated integrals:
Iterated integrals on ${\mathcal M}_{1,n}$ with integration along $d\tau$ and
iterated integrals on ${\mathcal M}_{1,n}$ with integration along the $dz_j$'s.

In addition we have to specify the differential one-forms we want to integrate.
The differential one-forms which we want to consider in the case of ${\mathcal M}_{1,n}$
are derived from the Kronecker function.
The Kronecker function $F(x,y,\tau)$ is defined in terms of the first Jacobi theta function by
\bq
 F\left(x,y,\tau\right)
 & = &
 \pi
 \theta_1'\left(0,q\right) \frac{\theta_1\left( \pi\left(x+y\right), q \right)}{\theta_1\left( \pi x, q \right)\theta_1\left( \pi y, q \right)},
\eq
where $q=\exp(\pi i \tau)$
and $\theta_1'$ denotes the derivative with respect to the first argument.
The first Jacobi theta function $\theta_1(z,q)$ is defined by
\bq
\theta_1\left(z,q\right) 
 & = &
 -i \sum\limits_{n=-\infty}^\infty \left(-1\right)^n q^{\left(n+\frac{1}{2}\right)^2} e^{i\left(2n+1\right)z},
 \;\;\;\;\;\;\;\;\;
 q = e^{i \pi \tau}.
\eq
Please note that in order to make contact with the standard notation for the Jacobi
theta functions we used here the nome $q=\exp(\pi i \tau)$ and not the nome squared
$\bar{q}=q^2=\exp(2 \pi i \tau)$.
The definition of the Kronecker function is cleaned up if we define
\bq
 \bar{\theta}_1\left(z,\bar{q}\right) 
 & = &
 \theta_1\left(\pi z,\bar{q}^{\frac{1}{2}} \right).
\eq
Then
\bq
 F\left(x,y,\tau\right)
 & = &
 \bar{\theta}_1'\left(0,\bar{q}\right) \frac{\bar{\theta}_1\left(x+y, \bar{q}\right)}{\bar{\theta}_1\left(x, \bar{q}\right)\bar{\theta}_1\left(y, \bar{q}\right)}.
\eq
It is obvious from the definition that the Kronecker function is symmetric in $x$ and $y$.
We are interested in the Laurent expansion in one of these variables.
We define functions
$g^{(k)}(z,\tau)$ through
\bq
\label{weinzierl:def_g_n}
 F\left(z,\alpha,\tau\right)
 & = &
 \sum\limits_{k=0}^\infty g^{(k)}\left(z,\tau\right) \alpha^{k-1}.
\eq
We are primarily interested in the coefficients $g^{(k)}(z,\tau)$ of the Kronecker function.
Let us recall some of their properties \cite{Zagier:1991,Brown:2011,Broedel:2018qkq}.
\begin{enumerate}

\item When viewed as a function of $z$, the function $g^{(k)}(z,\tau)$ has only simple poles.
More concretely, the function $g^{(1)}(z,\tau)$ has a simple pole with unit residue at every point of the lattice.
For $k>1$ the function $g^{(k)}(z,\tau)$ has a simple pole only at those lattice points 
that do not lie on the real axis.

\item The (quasi-) periodicity properties are
\bq
 g^{(k)}\left(z+1,\tau\right) & = &  g^{(k)}\left(z,\tau\right),
 \nonumber \\
 g^{(k)}\left(z+\tau,\tau\right) & = &  
 \sum\limits_{j=0}^k \frac{\left(-2\pi i\right)^j}{j!} g^{(k-j)}\left(z,\tau\right).
\eq
We see that $g^{(k)}(z,\tau)$ is invariant under translations by $1$, but not by $\tau$.

\item The functions $g^{(k)}(z,\tau)$ have the symmetry
\bq
 g^{(k)}(-z,\tau)
 & = &
 \left(-1\right)^k g^{(k)}(z,\tau).
\eq

\item Under modular transformations the functions $g^{(k)}(z,\tau)$ transform as
\bq
 g^{(k)}\left(\frac{z}{c\tau+d},\frac{a\tau+b}{c\tau+d}\right)
 & = &
 \left(c\tau +d \right)^k
 \sum\limits_{j=0}^k
 \frac{\left(2\pi i\right)^j}{j!}
 \left( \frac{c z}{c\tau+d} \right)^j
 g^{(k-j)}\left(z,\tau\right).
 \;\;
\eq

\item The $\bar{q}$-expansion of the $g^{(k)}(z,\tau)$ functions is given by 
(with $\bar{q}=\exp(2\pi i\tau)$ and $\bar{w}=\exp(2\pi i z)$)
\bq
 g^{(0)}\left(z,\tau\right)
 & = & 1,
 \nonumber \\
 g^{(1)}\left(z,\tau\right)
 & = &
 - 2 \pi i \left[
                  \frac{1+\bar{w}}{2 \left(1-\bar{w}\right)}
                  + \overline{\mathrm{E}}_{0,0}\left(\bar{w};1;\bar{q}\right)
 \right],
 \nonumber \\
 g^{(k)}\left(z,\tau\right)
 & = &
 - \frac{\left(2\pi i\right)^k}{\left(k-1\right)!} 
 \left[
 - \frac{B_k}{k}
       + \overline{\mathrm{E}}_{0,1-k}\left(\bar{w};1;\bar{q}\right)
 \right],
 \;\;\;\;\;\;\;\;\;\;\;\;\;\;\;\;\;\;\;\;\;\;\;\;\;\;\;
 k > 1,
\eq
where $B_k$ denote the $k$-th Bernoulli number, defined by
\bq
 \frac{x}{e^x-1}
 & = &
 \sum\limits_{k=0}^\infty \frac{B_k}{k!} x^k,
\eq
and
\bq
 \overline{\mathrm{E}}_{n;m}\left(\bar{u};\bar{v};\bar{q}\right) 
 & = &
  \mathrm{ELi}_{n;m}\left(\bar{u};\bar{v};\bar{q}\right)
  - \left(-1\right)^{n+m} \mathrm{ELi}_{n;m}\left(\bar{u}^{-1};\bar{v}^{-1};\bar{q}\right),
 \nonumber \\
 \mathrm{ELi}_{n;m}\left(\bar{u};\bar{v};\bar{q}\right) & = & 
 \sum\limits_{j=1}^\infty \sum\limits_{k=1}^\infty \; \frac{\bar{u}^j}{j^n} \frac{\bar{v}^k}{k^m} \bar{q}^{j k}.
\eq

\end{enumerate}
Having defined the functions $g^{(k)}(z,\tau)$, we may now state the differential one forms
which we would like to integrate on ${\mathcal M}_{1,n}$.
To keep the discussion simple, we focus on ${\mathcal M}_{1,2}$ with coordinates $(\tau,z)$.
(The general case ${\mathcal M}_{1,n}$ is only from a notational perspective more cumbersome.)
We consider
\bq
\label{weinzierl:omega_Kronecker}
 \omega^{\mathrm{Kronecker}}_{k}
 & = &
 \left(2\pi i\right)^{2-k}
 \left[
  g^{(k-1)}\left( z-c_j, \tau\right) d z + \left(k-1\right) g^{(k)}\left( z-c_j, \tau\right) \frac{d\tau}{2\pi i}
 \right],
 \;\;\;
\eq
with $c_j$ being a constant.
The differential one-form $\omega^{\mathrm{Kronecker}}_{k}$ is closed
\bq
 d \omega^{\mathrm{Kronecker}}_{k} & = & 0.
\eq
For the integration along $dz$ (i.e. $\tau=\mathrm{const}$) the part
\bq
\label{weinzierl:omega_Kronecker_dz}
 \omega^{\mathrm{Kronecker},z}_{k}
 & = &
 \left(2\pi i\right)^{2-k}
 g^{(k-1)}\left( z-c_j, \tau\right) d z
\eq
is relevant.
The iterated integrals of the differential one-forms in eq.~(\ref{weinzierl:omega_Kronecker_dz})
along a path $\gamma$ from $z=0$ to $z$ are the elliptic multiple polylogarithms $\widetilde{\Gamma}$,
as defined in ref.~\cite{Broedel:2017kkb}:
\bq
\label{weinzierl:Gammatilde}
\lefteqn{
 \widetilde{\Gamma}\!\left({\begin{smallmatrix} n_1 & ... & n_r \\ c_1 & ... & c_r \\ \end{smallmatrix}}; z; \tau \right)
 = } & &
 \nonumber \\
 & &
 \left( 2 \pi i\right)^{n_1+\dots+n_r-r}
 I_\gamma\left( \omega^{\mathrm{Kronecker},z}_{n_1+1}\left(c_1,\tau\right), \dots, \omega^{\mathrm{Kronecker},z}_{n_r+1}\left(c_r,\tau\right); z \right).
\eq
It is not possible that the differential one-forms $\omega$ entering the definition of elliptic multiple polylogarithms
have at the same time the following three properties:
(i) $\omega$ is double-periodic,
(ii) $\omega$ is meromorphic and
(iii) $\omega$ has only simple poles.
We can only require two of these three properties.
The definition of the $\widetilde{\Gamma}$-functions
selects meromorphicity and simple poles.
The differential one-forms are not double-periodic. 
(This is spoiled by the quasi-periodicity of $g^{(k)}( z, \tau)$ in $\tau$.)
However, this is what physics (i.e. the evaluation of Feynman integrals)
dictates us to choose.
The integrands are then either multi-valued functions on ${\mathcal M}_{1,n}$
or single-valued functions
on a covering space,
in the same way as $\ln(z)$ is a multi-valued function on ${\mathbb C}^\times$ or a single-valued function
on a covering space of ${\mathbb C}^\times$.
Of course, in mathematics one might also consider alternative definitions, which prioritise other properties.
A definition of elliptic multiple polylogarithms, which implements properties (i) and (ii), but gives up property (iii)
can be found in \cite{Levin:2007},
a definition, which implements properties (i) and (iii), but gives up (ii) can be found in \cite{Brown:2011}.
It is a little bit unfortunate that these different function are all named elliptic multiple polylogarithms.
The reader is advised to carefully check what is meant by the name ``elliptic multiple polylogarithm''.

Let us now consider the integration along $d\tau$ (i.e. $z=\mathrm{const}$).
Here, the part
\bq
\label{weinzierl:omega_Kronecker_dtau}
 \omega^{\mathrm{Kronecker},\tau}_{k}
 & = &
 \left(2\pi i\right)^{2-k}
 \left(k-1\right) g^{(k)}\left( z-c_j, \tau\right) \frac{d\tau}{2\pi i}
 \nonumber \\
 & = &
 \frac{\left(k-1\right)}{\left(2\pi i\right)^{k}} g^{(k)}\left(z-c_j, \tau\right) \frac{d\bar{q}}{\bar{q}}
\eq
is relevant.
This is supplemented by $z$-independent differential one-forms constructed from modular forms:
Let $f_k(\tau)$ be a modular form of weight $k$. We set
\bq
\label{weinzierl:omega_modular}
 \omega^{\mathrm{modular}}_{k}
 & = &
 \left( 2 \pi i \right) f_k\left(\tau\right) d\tau
 \;\; = \;\;
 f_k\left(\tau\right) \frac{d\bar{q}}{\bar{q}}.
\eq
Let $\omega_{k_j}$ be as in eq.~(\ref{weinzierl:omega_Kronecker_dtau}) or as in eq.~(\ref{weinzierl:omega_modular})
and $\gamma$ the path from $\tau=i\infty$ to $\tau$, corresponding in $\bar{q}$-space to a path from $\bar{q}=0$ to $\bar{q}$.
We then consider in $\bar{q}$-space the iterated integrals 
\bq
 I_\gamma\left( \omega_{k_1}, \dots, \omega_{k_r}; \bar{q} \right).
\eq
The integrands have no poles in $0 < | \bar{q} |< 1$.
A simple pole at $\bar{q}=0$ is possible and allowed.
If $\omega_{k_r}$ has simple pole $\bar{q}=0$ we say that the iterated integral has a trailing zero.
We may split $\omega_{k_r}$ into a part proportional to $d\bar{q}/\bar{q}$ and a regular remainder.
The singular part of a trailing zero can be treated in exactly the same way as we did in the case of multiple polylogarithms.


\section{Physics}
\label{weinzierl_section_physics}

After reviewing the mathematical background let us now turn to physics, and here in particular 
to the computation of Feynman integrals.

Integration-by-parts identities \cite{Tkachov:1981wb,Chetyrkin:1981qh}
and differential equations \cite{Kotikov:1990kg,Kotikov:1991pm,Remiddi:1997ny,Gehrmann:1999as,Argeri:2007up,MullerStach:2012mp,Henn:2013pwa,Henn:2014qga,Ablinger:2015tua,Adams:2017tga,Bosma:2017hrk}
are standard tools for the computation of Feynman integrals.
In essence, integration-by-parts identities allow us to express a Feynman integral
from a large set of Feynman integrals as a linear
combination of Feynman integrals from a smaller set. 
The Feynman integrals in the smaller set are called master integrals and
we may think of the master integrals as a basis of an (abstract) vector space.
We denote the number of master integrals by $N_F = N_{\mathrm{Fibre}}$
and the master integrals by $I = (I_1, ..., I_{N_F})$.
The notation used in relation with Feynman integrals is summarised in table~\ref{weinzierl:table2}.
\begin{table}
\begin{center}
\begin{tabular}{|lll|}
\hline
$N_F = N_{\mathrm{Fibre}}$: & Number of master integrals, & \\
 & master integrals denoted by & $I = (I_1, ..., I_{N_F})$. \\
 & & \\
$N_B = N_{\mathrm{Base}}$: & Number of kinematic variables, & \\
 & kinematic variables denoted by & $x=(x_1, ..., x_{N_B})$. \\ 
 & & \\
$N_L = N_{\mathrm{Letters}}$: & Number of letters, & \\
 & differential one-forms denoted by & $ \omega=(\omega_1, ..., \omega_{N_L})$. \\
\hline
\end{tabular}
\end{center}
\caption{
The notation used in connection with Feynman integrals:
$N_F$ denotes the number of master integrals, $N_B$ the number of kinematic variables the master integrals depend on
and $N_L$ the number linearly independent differential one-forms appearing in the $\eps$-form of the differential equation.
}
\label{weinzierl:table2}
\end{table}
Public available computer programs 
based on the Laporta algorithm \cite{Laporta:2001dd}
like 
\verb|REDUZE| \cite{vonManteuffel:2012np},
\verb|FIRE| \cite{Smirnov:2014hma} or
\verb|KIRA| \cite{Maierhoefer:2017hyi}
can be used to perform the reduction to the 
master integrals.

For the master integrals one derives (again by using integration-by-parts identities)
differential equations in the external invariants or internal masses.
We denote the number of kinematic variables by $N_B = N_{\mathrm{Base}}$ and the kinematic variables by $x=(x_1, ..., x_{N_B})$.
The system of differential equations for the master integrals can be written as
\bq
\label{weinzierl:dgl}
 \left( d + A \right) I & = & 0,
\eq
where $A(\eps,x)$ is a matrix-valued one-form
\bq
 A & = & 
 \sum\limits_{i=1}^{N_B} A_i dx_i.
\eq
The $A_i(\eps,x)$'s are matrices of size $N_F \times N_F$,
whose entries are rational functions in the dimensional regularisation parameter $\eps$
and the kinematic variables $x$.
The matrix-valued one-form $A$ satisfies the integrability condition
\bq
 dA + A \wedge A & = & 0.
\eq
Geometrically we have a vector bundle with a fibre of dimension $N_F$ spanned by $I_1, \dots I_{N_F}$ 
and a base space of dimension $N_B$ with local coordinates $x_1, \dots, x_{N_B}$.
The matrix-valued one-form $A$ defines a flat connection.

Up to this point everything is general and applies to any Feynman integral.
In particular, computing a Feynman integral is reduced to the problem of solving a system of differential equations as in eq.~(\ref{weinzierl:dgl}).
The solution of a system of differential equations requires in addition boundary values.
The boundary values correspond to simpler Feynman integrals,
where some kinematic variables have special values or vanish.
Therefore at this stage the boundary values can be considered to be known
(otherwise one would first set up a system of differential equations for the boundary values).

The system of differential equations is particular simple \cite{Henn:2013pwa}, if $A$ is of the form
\bq
\label{weinzierl:dgl_eps_form}
 A & = &
 \eps \; \; \sum\limits_{j=1}^{N_L} \; C_j \; \omega_j,
\eq
where
\begin{description}
\item{-} the only dependence on the dimensional regularisation parameter $\eps$ is given by the explicit prefactor,
\item{-} the $C_j$'s are $N_F \times N_F$-matrices, whose entries are numbers $r_1+ir_2$ with $r_1,r_2 \in \mathbb{Q}$,
\item{-} the differential one-forms $\omega_j$ have only simple poles (and depend only on $x$).
\end{description}
We denote by $N_L = N_{\mathrm{Letters}}$ the number of letters, i.e. the number of ${\mathbb Q}[i]$-linear independent differential
one-forms $\omega_j$.
The set of letters is denoted by $ \omega=(\omega_1, ..., \omega_{N_L})$.

Let us now discuss the possibilities to transform a generic system of differential equations
as in eq.~(\ref{weinzierl:dgl}) into the simple form of eq.~(\ref{weinzierl:dgl_eps_form}).
On the one hand we may
change the basis of the master integrals
\bq
 I' & = & U I,
\eq
where $U(\eps,x)$ is a $N_F \times N_F$-matrix.
The new connection matrix is
\bq
 A' & = & U A U^{-1} + U d U^{-1}.
\eq
On the other hand, we may perform a coordinate transformation on the base manifold:
\bq
 x_i' & = & f_i\left(x\right), \;\;\;\;\;\;\;\;\; 1 \le i \le N_B.
\eq
The connection transforms as 
\bq
 A \; = \; \sum\limits_{i=1}^{N_B} A_i dx_i
 & \;\;\;\;\;\; \Rightarrow \;\;\;\;\;\; &
 A' \; = \; \sum\limits_{i,j=1}^{N_B} A_i \; \frac{\partial x_i}{\partial x_j'} \; dx_j'.
\eq
Let us consider some examples of elliptic Feynman integrals.
The most prominent example is the two-loop sunrise integral.
The two-loop sunrise integral is defined by
\bq
\label{weinzierl:def_sunrise}
 S_{\nu_1 \nu_2 \nu_3}\left( \eps, x \right)
 =
 \left(-1\right)^{\nu_{123}}
 e^{2\gamma_E \eps}
 \left(m_3^2\right)^{\nu_{123}-D}
 \int \frac{d^Dk_1}{i \pi^{\frac{D}{2}}} \frac{d^Dk_2}{i \pi^{\frac{D}{2}}}
 \frac{1}{D_1^{\nu_1} D_2^{\nu_2} D_3^{\nu_3}},
\eq
with the propagators
\bq
\label{weinzierl:def_propagators}
 D_1=k_1^2-m_1^2,
 \hspace{0.3cm}  
 D_2 = (k_1-k_2)^2-m_2^2,
 \hspace{0.3cm} 
 D_3 = (p-k_2)^2-m_3^2
\eq
and $\nu_{123}=\nu_1+\nu_2+\nu_3$.
$\gamma_E$ denotes Euler's constant.
It is convenient to consider this Feynman integral in $D=2-2\eps$ space-time dimensions.
With the help of dimensional shift relations \cite{Tarasov:1996br,Tarasov:1997kx}
the result in $D=2-2\eps$ dimensions is easily related to
the corresponding Feynman integrals in $D=4-2\eps$ dimensions.

The simplest example for an elliptic Feynman integral is the equal mass sunrise integral.
In the equal mass case we have $m_1=m_2=m_3=m$.
In this case we have $3$ master integrals and one kinematic variable, which we may take 
originally as $x = p^2/m^2$.
This corresponds to $N_F=3$ and $N_B=1$. 
In mathematical terms we are looking at a rank $3$ vector bundle over ${\mathcal M}_{1,1}$.

The first question which we should address is how to obtain the elliptic curve associated to this integral.
For the sunrise integral there are two possibilities, 
we may either obtain an elliptic curve from the Feynman graph polynomial or from the maximal cut.
The sunrise integral has three propagators, hence we need three Feynman parameters, which we denote by
$\alpha_1,\alpha_2,\alpha_3$.
The second graph polynomial defines an elliptic curve
\bq
 E^{\mathrm{Feynman}}
 & : &
 - \alpha_1 \alpha_2 \alpha_3 x + \left( \alpha_1 + \alpha_2 + \alpha_3 \right) \left( \alpha_1 \alpha_2 + \alpha_2 \alpha_3 + \alpha_3 \alpha_1 \right) 
 \; = \; 0,
\eq
in $\mathbb{CP}^2$, with $[\alpha_1:\alpha_2:\alpha_3]$ being the homogeneous coordinates of $\mathbb{CP}^2$.
The elliptic curve varies with the kinematic variable $x$.
In general, the Feynman parameter space can be viewed as $\mathbb{CP}^{n-1}$, with $n$ being the number of propagators
of the Feynman integral.
It is clear that this approach does not generalise in a straightforward way to other elliptic Feynman integrals
with more than three propagators. (For an elliptic curve we want the zero set of a single polynomial in $\mathbb{CP}^2$).

We therefore turn to the second method of obtaining the elliptic curve, which generalises easily:
From the maximal cut of the sunrise integral we obtain the elliptic curve
as a quartic polynomial $P(w,z)=0$:
\bq
\label{weinzierl:def_elliptic_curve}
 E^{\mathrm{cut}}
 & : &
 w^2 - z
       \left(z + 4 \right) 
       \left[z^2 + 2 \left(1+x\right) z + \left(1-x\right)^2 \right]
 \; = \; 0.
\eq
Also this elliptic curve varies with the kinematic variable $x$.
Please note that these two elliptic curves $E^{\mathrm{Feynman}}$ and $E^{\mathrm{cut}}$ 
are not isomorphic, but only isogenic.
Let $\omega_1$ and $\omega_2$ be two periods of this elliptic curve $E^{\mathrm{cut}}$
with $\mathrm{Im}(\omega_2/\omega_1)>0$
and set $\tau=\omega_2/\omega_1$.
We denote the Wronskian by
\bq
\label{weinzierl:def_Wronskian}
 W & = & \omega_{1} \frac{d}{dx} \omega_{2} - \omega_{2} \frac{d}{dx} \omega_{1}.
\eq
In order to bring the system of differential equations for the equal mass sunrise integral into the simple form of eq.~(\ref{weinzierl:dgl_eps_form})
we perform a change of the basis of the master integrals from a pre-canonical basis $(S_{110},S_{111},S_{211})$ to
\bq
\label{weinzierl:def_basis}
 J_1
 & = &
 4 \eps^2 \; S_{110}\left(\eps,x\right),
 \nonumber \\
 J_2
 & = &
 \eps^2 \frac{\pi}{\omega_1} \; S_{111}\left(\eps,x\right),
 \nonumber \\
 J_3
 & = &
 \frac{1}{\eps} \frac{\omega_1^2}{2 \pi i W} \frac{d}{dx} J_2 
 + \frac{\omega_1^2}{2 \pi i W} \frac{\left(3x^2-10x-9\right)}{2x\left(x-1\right)\left(x-9\right)} J_2.
\eq
This transformation is not rational or algebraic in $x$, as can be seen from the prefactor $1/\omega_1$ in the definition of $J_2$.
The period $\omega_1$ is a transcendental function of $x$.
In addition we change the kinematic variable from $x$ to $\tau$ (or $\bar{q}$).
Again, this is a non-algebraic change of variables.
One obtains
\bq
\left( d + A \right) J & = & 0
\eq
with
\bq
 A & = &
 2 \pi i \; \eps
 \left( \begin{array}{ccc}
 0 & 0 & 0 \\
 0 & \eta_2\left(\tau\right) & \eta_0\left(\tau\right) \\
 \eta_3\left(\tau\right) & \eta_4\left(\tau\right) & \eta_2\left(\tau\right) \\
 \end{array} \right)
 d\tau,
\eq
where $\eta_k(\tau)$ denotes a modular form of modular weight $k$ for $\Gamma(6)$.
The differential equation for the equal mass sunrise system is now in $\eps$-form and the kinematic variable
matches the standard coordinate on ${\mathcal M}_{1,1}$.
With the additional information of a boundary value, the differential equation is now easily solved order by order
in $\eps$ in terms of iterated integrals of modular forms.
One finds for example
\bq
 J_2 & = &
 \left[ 3 \, \mathrm{Cl}_2\left(\frac{2\pi}{3}\right) 
 +I\left(\eta_0,\eta_3;\tau\right)
 \right] \eps^2
 + {\mathcal O}\left(\eps^3\right).
\eq
The Clausen value $\mathrm{Cl}_2(2\pi/3)$ comes from the boundary value.

Let us also consider an example where the kinematic space is ${\mathcal M}_{1,n}$ with $n>1$.
We don't have to go very far, the unequal mass sunrise integral provides an example.
We now take the three masses squared $m_1^2$, $m_2^2$ and $m_3^2$ in eq.~(\ref{weinzierl:def_propagators})
to be pairwise distinct.
We now have $7$ master integrals and $3$ kinematic variables.
As original kinematic variables we use
$x = p^2/m_3^2$,
$y_1 = m_1^2/m_3^2$,
$y_2 = m_2^2/m_3^2$.
This corresponds to $N_F=7$ and $N_B=3$. 
In mathematical terms we are looking at a rank $7$ vector bundle over ${\mathcal M}_{1,3}$.

Finding the elliptic curve proceeds exactly in the same way as discussed in the equal mass case.
In the next step we would like to change the kinematic variables from $(x,y_1,y_2)$ to the standard
coordinates $(\tau,z_1,z_2)$ on ${\mathcal M}_{1,3}$.
This raises the question: How to express the new coordinates in terms of the old coordinates (or vice versa)?
For $\tau$ the answer is straightforward: $\tau$ is a again the ratio of the two periods $\tau=\omega_2/\omega_1$,
and $\omega_1$ and $\omega_2$ are functions of $x$, $y_1$ and $y_2$.

Also for $z_1$ and $z_2$ there is a simple geometric interpretation:
In the Feynman parameter representation there are two geometric objects of interest:
the domain of integration $\sigma$ (the simplex $\alpha_1, \alpha_2, \alpha_3 \ge 0$, $\alpha_1+\alpha_2\alpha_3\le 1$)
and the elliptic curve $E^{\mathrm{Feynman}}$ (the zero set $X$ of the second graph polynomial).
\begin{figure}
\begin{center}
\includegraphics[scale=0.9]{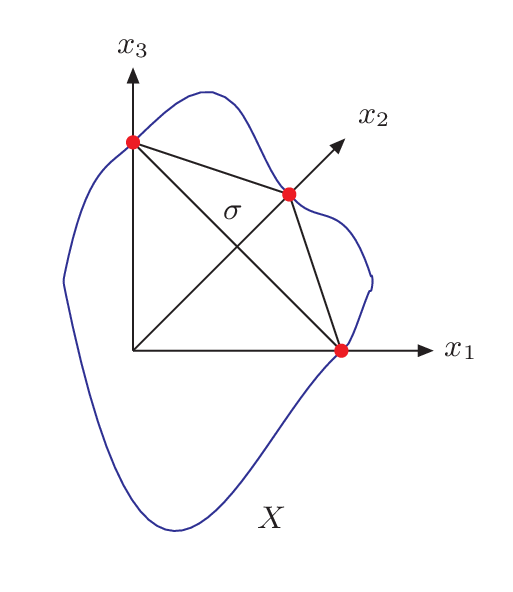}
\includegraphics[scale=0.9]{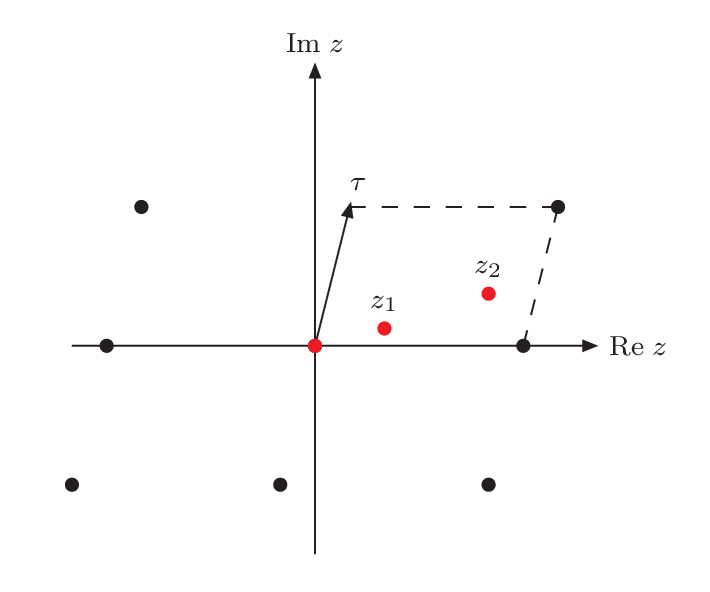}
\end{center}
\caption{
$X$ and $\sigma$ intersect at three points, the images of these three points in ${\mathbb C}/\Lambda$ are $0,z_1,z_2$.
}
\label{weinzierl:fig6}
\end{figure}
$X$ and $\sigma$ intersect at three points, as shown in fig.~\ref{weinzierl:fig6}.
The images of these three points in ${\mathbb C}/\Lambda$ are $0,z_1,z_2$, where we used a translation transformation to fix one point at $0$.

The system of differential equations can again be transformed into the simple form of eq.~(\ref{weinzierl:dgl_eps_form}) by a redefinition of the master integrals 
and a change of coordinates from $(x,y_1,y_2)$ to $(\tau,z_1,z_2)$.
The explicit formula for the fibre transformation is a little bit lengthy 
and can be found in the literature \cite{Bogner:2019lfa,Weinzierl:2020fyx}.
Doing so, one finds
\bq
\label{weinzierl:dgl_unequal_sunrise}
 A & = &
 \eps \; \; \sum\limits_{j=1}^{N_L} \; C_j \; \omega_j,
 \;\;\;\;\;\;\;\;\;\;\;\; 
 \mbox{with $\omega_j$ having only simple poles}, 
\eq
where $\omega_j$ is either 
\bq
 2\pi i \;
 f_k\left(\tau\right) d\tau,
\eq
where $f_k(\tau)$ is a modular form, or of the form
\bq
 \lefteqn{
 \omega_k\left( L\left(z\right), \tau\right)
 = } & &
 \nonumber \\
 & &
 \left(2\pi i\right)^{2-k}
 \left[
  g^{(k-1)}\left( L\left(z\right), \tau\right) d L\left(z\right) + \left(k-1\right) g^{(k)}\left( L\left(z\right), \tau\right) \frac{d\tau}{2\pi i}
 \right],
\eq
with 
$L(z)$ being a linear function of $z_1$ and $z_2$:
\bq
 L\left(z\right)
 & = &
 \sum\limits_{j=1}^{2} \alpha_j z_j + \beta,
\eq
and $\alpha_1$, $\alpha_2$ and $\beta$ being constants.

With the additional information of a boundary value, the differential equation 
in eq.~(\ref{weinzierl:dgl_unequal_sunrise})
is now easily solved order by order
in $\eps$ in terms of iterated integrals as discussed in section~\ref{weinzierl_section_moduli_spaces}.
We are free to choose a suitable point in kinematic space for the boundary value and to integrate the differential
equation from the chosen boundary point to the kinematic point of interest.
We are free to choose any path (as long as the path avoids branch cuts).
An arbitrary path will involve integrations along $d\tau$ and the $dz_j$'s.
It is advantageous to use as boundary condition the values on the hypersurface $\tau=i\infty$.
There the elliptic curve degenerates, i.e. the geometric genus equals zero,
and the sought-after boundary values of the Feynman integrals are expressible in terms of multiple polylogarithms.
We may then integrate the differential equation only along $d\tau$.
This avoids integrations along the $dz_j$'s, the
analytic expressions tend to be shorter and
the final expressions are easier to evaluate numerically.
This approach also avoids poles and branch cuts along the integration path.
The only problem, which might occur, are a slow convergence of the $\bar{q}$-expansion of the final result
in the case $|\bar{q}| \lesssim 1$.
This can be avoided by using in this kinematic region a different choice of periods $\omega_1$ and $\omega_2$,
related to the original ones by a modular transformation \cite{Weinzierl:2020fyx}.
It is therefore always possible to achieve
\bq
 \left| \bar{q} \right|
 & \le &
 e^{- \pi \sqrt{3}}
 \; \approx \; 0.0043,
\eq
which guarantees a fast convergence of the $\bar{q}$-expansion of the final result.


\section{Conclusions}
\label{weinzierl_section_conclusions}

Feynman integrals are important in many areas of physics and indispensable for precision calculations
within perturbation theory beyond the leading order.
Feynman integrals, which evaluate to multiple polylogarithms are by now well understood.
Multiple polylogarithms are iterated integrals on the moduli space ${\mathcal M}_{0,n}$.
From two loops onwards, there is a class of Feynman integrals related to elliptic curves,
which evaluate to iterated integrals on the moduli space ${\mathcal M}_{1,n}$.
These integrals were the main topic of this talk.
We discussed the mathematical background of elliptic curves, elliptic functions, modular forms and the moduli space
of $n$ marked points on a smooth curve of genus one.
The investment in the mathematical foundations pays off, as with the right language we may transfer methods
known from the genus zero case to the genus one case.
In particular we may achieve through a redefinition of the master integrals and a change of the kinematic variables
that the differential equation is transformed to
\bq
 A & = &
 \eps \; \; \sum\limits_{k=1}^{N_L} \; C_k \; \omega_k,
 \;\;\;\;\;\;\;\;\;\;\;\; 
 \mbox{with $\omega_k$ having only simple poles}. 
\eq
This form can be reached for many Feynman integrals evaluating to multiple polylogarithms and
-- as we have seen in this talk --
also for non-trivial elliptic Feynman integrals.

{\footnotesize
\bibliography{/home/stefanw/notes/biblio}
\bibliographystyle{/home/stefanw/latex-style/h-physrev5}
}

\end{document}